# On the complexity of inducing categorical and quantitative association rules

Fabrizio Angiulli
ISI-CNR
c/o DEIS, Università della Calabria,
Via Pietro Bucci, 41C
87036 Rende (CS), Italy
E-mail: angiulli@isi.cs.cnr.it
Phone: +39 0984 831738
Fax: +39 0984 839054

Giovambattista Ianni
DEIS, Università della Calabria,
Via Pietro Bucci, 41C
87036 Rende (CS), Italy
E-mail: ianni@deis.unical.it
Phone: +39 0984 494749
Fax: +39 0984 494713

Luigi Palopoli
DIMET, Università di Reggio Calabria,
Via Graziella, Loc. Feo di Vito
89100 Reggio Calabria (RC), Italy
E-mail: palopoli@ing.unirc.it
Phone: +39 0965 875235
Fax: +39 0965 875481

**Abstract**

Inducing association rules is one of the central tasks in data mining applications. Quantitative association rules induced from databases describe rich and hidden relationships holding within data that can prove useful for various application purposes (e.g., market basket analysis, customer profiling, and others). Even though such association rules are quite widely used in practice, a thorough analysis of the computational complexity of inducing them is missing. This paper intends to provide a contribution in this setting. To this end, we first formally define quantitative association rule mining problems, which entail boolean association rules as a special case, and then analyze their computational complexities, by considering both the standard cases, and a some special interesting case, that is, association rule induction over databases with null values, fixed-size attribute set databases, sparse databases, fixed threshold problems.

## 1 Introduction

The enormous growth of information available in database systems has pushed a significant development of techniques for knowledge discovery in databases. At the heart of the knowledge discovery process there is the application of data mining algorithms that are in charge of extracting hidden relationships holding among pieces of information stored in a given database [9]. Most used data mining algorithms include classification techniques, clustering analysis and association rule induction [2]. In this paper, we focus on this latter data mining technique. Informally speaking, an association rule tells that a conjunction of conditions implies a consequence. For instance, the rule *hamburger, fries* $\Rightarrow$ *soft−drink* induced from a purchase database, tells that a customer purchasing hamburgers and fries also purchases a soft-drink.

An association rule induced from a database is interesting if it describes a relationship that is, in a sense, "valid" as far as the information stored in the database is concerned. To state such validity, *indices* are used, that are functions with values usually in $[0, 1]$, that tell to what extent an extracted association rule describe knowledge valid in the database at hand. For instance a *confidence* value of 0.7 associated to the rule above tells that 70 percent of purchases including hamburgers and fries also include a soft-drink. In the literature, several index definitions have been provided (see e.g. [7], where many interestingness criteria are proposed). Clear enough, information patterns expressed in the form of association rules and associated indices indeed denote knowledge that can be useful in several application contexts, e.g., market basket analysis.

In some application contexts, however, *Boolean* association rules, like the one above are not expressive enough for the purposes of the given knowledge discovery task. In order to obtain more expressive association rules, one can allow more general forms of conditions to occur therein. *Quantitative association rules* [17] are ones where both the premise and the consequent use conditions of one of the following forms: $(i)$ $A = u$; $(ii)$ $A \neq u$; $(iii)$ $A' \in [l', u']$; $(iv)$ $A' \notin [l', u']$, where $A$ is a *categorical* attribute, i.e., an attribute that has associated a discrete, unordered domain and $u$ is a value in this domain, and $A'$ is a *numeric* attribute, that is, one associated with an ordered domain of numbers, and $l'$ and $u'$ ($l' \leq u'$) are two, not necessarily distinct, values. For instance, the quantitative rule

$(hamburger \in [2, 4]), (ice\text{-}cream\text{-}taste = chocolate) \Rightarrow (soft\text{-}drink \in [1, 3])$

induced from a purchase database, tells that a customer purchasing from 2 to 4 hamburgers and a chocolate ice-cream also purchases from 1 to 3 soft-drinks.

In either of their forms, inducing association rules is a quite widely used data mining technique, several systems have been developed based on them [3, 13], and several successful applications in various contexts have been described [8]. Despite the wide-spread utilization of association rule induction in practical applications, a thorough analysis of the complexity of the associated computational tasks have not been developed. However, such an analysis appears to be important since, as in other contexts, an appropriate understanding of the computational characteristics of the problem at hand makes it possible to single out tractable cases of generally untractable problems, isolate hard complexity sources and, overall, to devise more effective approaches to algorithm development.

As far as we know, some computational complexity analysis pertaining association rules are performed in [11, 14, 15, 19, 20]. In [14] and [15], a NP-hardness result is stated regarding the induction of association rules (or, in general, of *conditions*) having an optimal *entropy* (resp. *chi-square*); in [19], under some restrictive assumptions, the NP-completeness of inducing quantitative association rules with a *confidence* and a *support*[1] greater than two given thresholds is proved along with a result stating a polynomial bound on the complexity of mining quantitative rules over databases where the number of possible items is constant. In [11], it is stated the $\#P$-hardness of counting the number of mined association rules (under support measure), and moreover, a specialization of the result stated in Theorem 3.1 below regarding boolean association rules. Furthermore, [20] gives some results about the computational complexity of mining frequent itemsets under combined constraints on the number of items and on the frequency threshold.

In this paper we define a generalized form of association rules embracing both the quantitative and the categorical and the boolean types, in which null values (in the following indicated by $\epsilon$) denoting the absence of information, are used.

Nulls are often useful in practice. As an example, consider a market database in which attributes correspond to available products and values represent quantities sold. Null values can be used to denote the absence of a product in a particular transaction (this is quite different than specifying the value 0 instead). As a further example, consider unavailable values in medical records representing clinical cases in analysis of patient

---

[1]Entropy, confidence and support are indices (see below).

data. We call a database allowing null values, a *database with nulls*.

When we induce association rules from databases with nulls, we require that conditions on attributes assuming the null value are always unsatisfied, i.e. that it is not possible to specify conditions on null values. A boolean association rule can be thus regarded as a special case of quantitative or categorical association rule mined on a database with nulls.

In this paper, we analyze the computational complexity implied by inducing association rules using four of the mostly used rule quality indices, namely, confidence, support, $\theta$-gain and $h$-laplace [7, 2]. In particular, we shall show that, in the standard case, and depending on the chosen index of reference, the complexity of the problem is either P or NP-complete. When databases with nulls are considered, independently of the reference index, the rule induction task is NP-complete.

Despite these negative results, there are many cases where the problem turns out to be very easy to compute: whenever the instance database is sparse (i.e. each transaction/tuple is very small with respect to the set of possible attributes), or when the attribute set at hand has constant size, for any index, we are able to show that the computational complexity of the rule induction problem is L; furthermore introducing some constraint on the input instance leads to problems with very low complexity such as $TC^0$ or $AC_2^0$. Problems with this kind of complexity are very efficiently parallelizable (recall that $AC_2^0 \subseteq TC^0 \subseteq NC^1$, whereas $L \subseteq NC^2$).

The plan of the paper is as follows. In the following section we give preliminary definitions. In Section 3 we state general complexity results about inducing association rules. Sparse databases and Fixed-schema complexity of rule induction are dealt with in Section 4 and 5 respectively. Finally, Section 6 collects an interesting set of special tractable cases.

## 2 Preliminaries

We begin by defining several concepts that will be used throughout the paper, including, among others, those of association rule induction problems and indices.

**Definition 2.1** An *attribute* is an identifier with an associate domain. A *categorical attribute* (resp., *numeric attribute*) is one whose domain is an unordered set of values (resp., a set of integer or rational numbers). Both categorical and numeric attributes include in their domain the special value $\epsilon$.

Let $A$ be an attribute. We denote by $\mathbf{dom}(A)$ the domain of $A$.

Let $A$ be a categorical or numerical attribute. We say that $A$ is *boolean* if $\mathbf{dom}(A) = \{\epsilon, c(A)\}$, where $c(A)$ denotes an arbitrary constant associated to $A$.

**Definition 2.2** Let $I$ be a set of attributes. A *database* $T$ on $I$ is a relation with duplicates having $I$ as set of attributes. Let $A \in I$ and let $t$ be a tuple of $T$. We denote by $t[A]$ the value of the attribute $A$ in the tuple $t$. The *size* $|t|$ of $t \in T$ is $|\{A \in I \mid t[A] \neq \epsilon\}|$. We denote by $\mathbf{dom}(A, T)$ the set $\{t[A] \mid t \in T\} - \{\epsilon\}$.

**Definition 2.3** Let $I$ be a set of attributes, and let $T$ be a database on $I$. We say that $T$ is a *database without nulls* if, for each $t \in T$, $|t| = |I|$. Otherwise we say that $T$ is a *database with nulls*.

**Definition 2.4** Given a database $T$ defined on a set of attributes $I$ we call $m_T$ the longest tuple in it. We say that $T$ is a *boolean database* if every attribute $A \in I$ is boolean.

A family $S$ of boolean databases is sparse if, for any $T \in S$, $|m_T|$ is $\mathcal{O}(\log |I|)$ where $I$ is the set of attributes which $T$ is defined on. Given a family $S$ of sparse databases, we will call *sparse database* each element $T \in S$.

**Definition 2.5** An *atomic condition* on $A$ is:

- an expression of the form $A = u$ or $A \neq u$, where $A$ is a categorical attribute and $u$ is a value in the domain of $A$ distinct from the $\epsilon$ value, or
- an expression of the form $A \in [l, u]$ or $A \notin [l, u]$, where $A$ is a numeric attribute and $l$ and $u$ ($l \leq u$) are two, not necessarily distinct, numeric values.

Whenever numerical attributes are involved, the notation $A = u$ (resp. $A \neq u$) can be regarded as syntactic shortcut for $A \in [u, u]$ (resp. $A \notin [u, u]$).

**Definition 2.6** Given a categorical attribute $A$, and a database $T$, we denote:
$\mathbf{dom}(A = u, T)$ (resp. $\mathbf{dom}(A \neq u, T)$) as the set $\mathbf{dom}(A, T) \cap \{u\}$ (resp. $\mathbf{dom}(A, T) - \{u\}$).
Let $A$ be a numerical attribute, we denote by $\mathbf{dom}(A \in [l, u], T)$ (resp. $\mathbf{dom}(A \notin [l, u], T)$) the set $\mathbf{dom}(A, T) \cap \mathcal{I}(A, [l, u])$ (resp. $\mathbf{dom}(A, T) - \mathcal{I}(A, [l, u])$), where $\mathcal{I}(A, [l, u])$ is the set $\{x \in \mathbf{dom}(A) \mid l \leq x \leq u\}$.

**Definition 2.7** A *condition* $C$ on a set of distinct attributes $A_1, \ldots, A_n$ is an expression of the form $C = C_1 \wedge \ldots \wedge C_n$, where each $C_i$ is an atomic condition on $A_i$, for each $i = 1, \ldots, n$. We denote by $\mathbf{att}(C)$ the set $A_1, \ldots, A_n$. The *size* $|C|$ of $C$ is $n$.

We are now in the condition of defining association rules and their semantics.

**Definition 2.8** Let $I$ be a set of attributes. An *association rule* on $I$ is an expression of the form $B \Rightarrow H$, where $B$ and $H$, called *body* and *head* of the rule resp., are two conditions on the sets of attributes $I_B$ and $I_H$ resp., such that $\emptyset \subset I_B, I_H \subset I$, and $I_B \cap I_H = \emptyset$. The *size* $|B \Rightarrow H|$ of the rule is $|B| + |H|$.

**Definition 2.9** Let $I$ be a set of attributes, let $T$ be a database on $I$, and let $t$ be a tuple of $T$. Let $A \in I$, and let $C_a$ be an atomic condition on $A$, we say that $t$ *satisfies* $C_a$, written $t \vdash C_a$, iff $t[A] \in \mathbf{dom}(C_a, T)$. Let $C = C_1 \wedge \ldots \wedge C_n$ be a condition, we say that $t$ *satisfies* $C$, written $t \vdash C$, iff $t \vdash C_i$, for each $i = 1, \ldots, n$. Otherwise we say that $t$ *does not satisfy* $C$, written $t \nvdash C$. By $T_C$ we denote the set of tuples $\{t \in T \mid t \vdash C\}$.

**Definition 2.10** Let $I$ be a set of attributes, and let $T$ be a database on $I$, and let $C$ be a condition on a subset on $I$. We say that $C$ is *trivial* if it contains at least an atomic condition $C_a$ such that $T_{C_a} = T$. Let $B \Rightarrow H$ be an association rule on $I$. We say that $B \Rightarrow H$ is *trivial* if $B \wedge H$ is trivial.

Trivial rules with suitable value of interest can be easily built. Thus, we will focus, in the following, our attention on non-trivial association rules.

When inducing association rules from databases in data mining applications, one is usually interested in obtaining rules that describe knowledge "largely" valid in the given database. This concept is captured by several notions of *indices*, which have been defined in the literature. In the following, we shall consider the most widely used of them, whose definitions are given next.

**Definition 2.11** Let $I$ be a set of attributes, let $T$ be a database on $I$, and let $B \Rightarrow H$ be an association rule on $I$. Then:

1. the *support* of $B \Rightarrow H$ in $T$, written $sup(B \Rightarrow H, T)$, is $\frac{|T_{B \wedge H}|}{|T|}$;

2. the *confidence* of $B \Rightarrow H$ in $T$, written $cnf(B \Rightarrow H, T)$, is $\frac{|T_{B \wedge H}|}{|T_B|}$;

3. Let $\theta$ be a rational number, $0 < \theta \leq 1$, then the $\theta$-*gain* of $B \Rightarrow H$ in $T$, written $gain_\theta(B \Rightarrow H, T)$, is $\frac{|T_{B \wedge H}| - \theta \cdot |T_B|}{|T|}$;

4. Let $h$ be a natural, $h \geq 2$, then the $h$-*laplace* of $B \Rightarrow H$ in $T$, written $laplace_h(B \Rightarrow H, T)$, is $\frac{|T_{B \wedge H}| + 1}{|T_B| + h}$.

Now that we have defined association rules and associated indices (that, in different forms, measure the validity of an association rule w.r.t. a database where it has been induced from), we are in the condition to formally define next the association rule induction problems.

**Definition 2.12** Let $I$ be a set of attributes, let $T$ be a database on $I$, let $k$, $1 \leq k \leq |I|$, be a natural number, and let $s$, $0 < s \leq 1$, be a rational number. Furthermore, let $\rho \in \{sup, cnf, laplace_h, gain_\theta\}$. The association rule induction problem $\langle I, T, \rho, k, s \rangle$ is as follows: *Is there a non-trivial association rule $R$ such that $|R| \geq k$ and $\rho(R, T) \geq s$?*

In general, we shall thus measure the complexity of association rule induction problems for the various index forms we have defined above. As a special case, we shall also consider the complexity of the induction problems when the attribute set $I$ is assumed to be not part of the input, in which case we will talk about *fixed schema complexity* of the association rule induction problem.

**Remark.** In the literature it is usually assumed that, in answering an association rule induction problem, one looks for rules which match some bounds in terms of two or more indices [7]. Here we preferred to split the problem as to refer to one index at a time. Indeed, this allows us to single out more precisely complexity sources, and, moreover, complexity measures for problems involving more than one index can be obtained fairly easily from problems involving only one index.

## 2.1 Complexity Classes

We assume the reader is familiar with basic concepts regarding computational complexity and, in particular, the complexity classes P (the decision problems solved by polynomial-time bounded deterministic Turing machines), NP (the decision problems solved by polynomial-time bounded non-deterministic Turing machines) and L (the decision problems solved by logspace-bounded deterministic Turing machines).

**Definition 2.13** MAJORITY gates are unbounded fan-in gates (with binary input and output) that output 1 if and only if more than half of their inputs are non-zero.

**Definition 2.14** A family $\{C_i\}$ of boolean circuits, s.t. $C_i$ accepts strings of size $i$, is uniform if there exists a Turing machine $\mathcal{T}$ which on input $i$ produces the circuit $C_i$. $\{C_i\}$ is said to be *logspace uniform* if $\mathcal{T}$ carries out its work using $O(\log i)$ space. Define $AC^0$ (resp. $TC^0$) as the class of decision problems solved by uniform families of circuits of polynomial size and constant depth, with AND, OR, and NOT (resp. MAJORITY and NOT) gates of unbounded fan-in [1, 6, 16].

**Definition 2.15** For any $k > 0$, $\#\mathrm{AC}^0_k$ is the class of functions $f : \{0,1\}^* \to \mathbf{N}$ computed by depth $k$, polynomial size uniform families of circuits with $+, \times$-gates (the usual arithmetic sum and product in $\mathbf{N}$) having unbounded fan-in, where each value incoming into the circuit can be either constant (where the allowed constant values are 1 and 0) or being an input value in the form $x_i$ or $1 - x_i$ (where the allowed input values are 1 and 0). Let $\#\mathrm{AC}^0 = \bigcup_{k>0} \#\mathrm{AC}^0_k$ [1].

Thus, $\#AC^0$ circuits accept the values 1 and 0 as inputs, but they are considered as natural numbers.

**Definition 2.16** $\mathrm{GapAC}^0$ is the class of all functions $f : \{0,1\}^* \to \mathbf{N}$ that can be expressed as the difference of two functions in $\#\mathrm{AC}^0$ [1, 5]. $\mathrm{PAC}^0$ is the class of languages $\{A \mid \exists f \in \mathrm{GapAC}^0, x \in A \iff f(x) > 0\}$ [1].

## 3 General complexity results

Here we investigate the complexity of evaluating $\langle I, T, \rho, k, s \rangle$ when $I, T, k$ and $s$ are all taken as input values.

**Definition 3.1** Let $I$ be a set of numerical attributes, and let $T$ be a database on $I$. Let $A$ be an attribute in $I$, and let $u$ be a value. Define

- $\mathbf{lub}(u, A, T) = \min\{v \in \mathbf{dom}(A, T) \mid v \geq u\}$, and
- $\mathbf{glb}(u, A, T) = \max\{v \in \mathbf{dom}(A, T) \mid v \leq u\}$.

Let $C = A \in [l, u]$ (resp. $C = A \notin [l, u]$) be a non trivial atomic condition such that $|T_C| > 0$. Define

$$\begin{aligned}\mathbf{bot}(C, T) &= A \in [\mathbf{lub}(l, A, T), \mathbf{glb}(u, A, T)] \\ &\quad (A \notin [\mathbf{lub}(l, A, T), \mathbf{glb}(u, A, T)] \text{ resp.})\end{aligned}$$

Let $C = C_1 \wedge \ldots \wedge C_n$ be a non trivial condition such that $|T_C| > 0$. Define

$$\mathbf{bot}(C, T) = \mathbf{bot}(C_1, T) \wedge \ldots \wedge \mathbf{bot}(C_n, T)$$

**Proposition 3.1** *Let $I$ be a set of numerical attributes, Let $T$ be a database on $I$, and let $C$ be a non trivial condition on a subset of $I$ such that $|T_C| > 0$. Then $T_C = T_{\mathbf{bot}(C,T)}$.*

**Proof.** Straightforward. □

Proposition 3.1 has the technically important consequence that we can restrict our attention to conditions and association rules including only values from the database of interest.

Now we prove that, when support is assumed as the reference index, the association rule mining problem is NP-complete both in presence or absence of nulls. We point out that the following result extends the two more specific results presented in [19], that applies only to boolean databases (there called *0/1-relations*), and in [11], that applies only to numerical databases without nulls and to conditions on intervals containing at least two distinct numbers.

**Proposition 3.2** *Consider the problem $\mathcal{P} = \langle I, T, sup, k, s \rangle$. If there exists a rule $B \Rightarrow H$ that is a solution for $\mathcal{P}$, then for each $k'$, $1 < k' \leq k$, there exists a rule $B' \Rightarrow H'$ of size $k'$ such that $sup(B' \Rightarrow H', T) \geq s$.*

**Theorem 3.1** *Given a database $T$ without nulls, the problem $\langle I, T, sup, k, s \rangle$ is NP-complete.*

**Proof.** (*Hardness*) The proof is by reduction of the problem CLIQUE, which is well-known to be NP-complete [10]. Let $G = (V, E)$ be an undirected graph, with set of nodes $V = \{v_1, \ldots, v_n\}$ and set of edges $E = \{e_1 = \{v_{p_1}, v_{q_1}\}, \ldots, e_m = \{v_{p_m}, v_{q_m}\}\}$. Let $h$ be an integer. The CLIQUE problem is: *Does there exist in $G$ a complete subgraph (clique) of size at least $h$?*

W.l.o.g. suppose the graph $G$ is connected. We build an instance of $\langle I, T, sup, k, s \rangle$ as follows: let $I^{clq}$ be the set consisting of the attributes $I_1, \ldots, I_n$, so that $I_j$ represents the node $v_j$ of $G$, for each $j = 1, \ldots, n$. Let $T^{clq}$ be the database on $I^{clq}$ formed by a tuple $t_{e_i}$, for each $i = 1, \ldots, m$, such that $t_{e_i}[I_j] = 0$ if $v_j \in e_i$, and $t_{e_i}[I_j] = 1$ otherwise ($t_{e_i}$ encodes the edge $e_i$ of $G$). Next, we prove that $G$ has a clique of size $k$ in $G$ iff $\langle I^{clq}, T^{clq}, sup, n-k, \frac{k(k-1)}{2m} \rangle$ is a YES instance.
We have the following fact.

**Fact 3.1** *Let $J \in I^{clq}$, let $C' = (I_j = 0)$ (or, equivalently $C' = (I_j \neq 1)$), and let $C''$ be a non trivial condition defined on a subset of $I^{clq} - \{I_j\}$. Then $|T^{clq}_{C' \wedge C''}| \leq n - |C' \wedge C''|$.*

We can resume Theorem's proof.
($\Rightarrow$) Let $C = \{v_{r_1}, \ldots, v_{r_k}\}$ be a clique of size $k$ in $G$. Consider the condition

$$B \wedge H = \left( \bigwedge_{v_j \in (V-C)} (I_j = 1) \right)$$

Since $G$ is connected, $B \wedge H$ is non trivial. By definition of clique, there exist $\frac{k(k-1)}{2}$ edges of $G$ connecting nodes in $C$. Therefore, the cardinality of

$$T' = \{t_{\{v_{r_x}, v_{r_y}\}} \in T^{clq} \mid 1 \leq x < y \leq k\}$$

equals $\frac{k(k-1)}{2}$. Clearly $T' \subseteq T^{clq}_{B \wedge H}$ and $sup(B \Rightarrow H, T^{clq}) \geq \frac{k(k-1)}{2m}$.

($\Leftarrow$) By Proposition 3.2, if $\langle I^{clq}, T^{clq}, sup, n-k, \frac{k(k-1)}{2m} \rangle$ is a YES instance then there exists a non trivial rule $B \Rightarrow H$ of size $n - k$ such that $|T^{clq}_{B \wedge H}| \geq \frac{k(k-1)}{2}$.
First, we note that atomic conditions on numerical attributes of the form $I_j \in [0, 1]$ are trivial, while the same does not apply to categorical attributes. W.l.o.g. assume $k \geq 4$. By contradiction, suppose that there exists a condition $I_j = 0$ (or $I_j \neq 1$) occurring in $B \Rightarrow H$, then, by Fact 3.1, $|T^{clq}_{B \wedge H}| \leq k < \frac{k(k-1)}{2}$. Hence only conditions of the form $I_j = 1$ ($I_j \neq 0$) can appear in $B \Rightarrow H$.
Let $I^{clq} - \mathbf{att}(B \wedge H) = \{I_{r_1}, \ldots, I_{r_k}\}$. In order to be $|T^{clq}_{B \wedge H}| \geq \frac{k(k-1)}{2}$, $T^{clq}_{B \wedge H}$ contains, at least, the set

$$\{t_{\{v_{r_x}, v_{r_y}\}} \in T^{clq} \mid 1 \leq x < y \leq k\}$$

i.e. the nodes $v_{r_1}, \ldots, v_{r_k}$ form a clique of $G$ having size $k$.
(*Membership*) Certificate: an association rule $B \Rightarrow H$ on a subset of $I$. Polynomial checking: verify that $B \Rightarrow H$ is non trivial, that $|B \Rightarrow H| \geq k$, and that $sup(B \Rightarrow H, T) \geq s$. □

**Theorem 3.2** *Given a database $T$ with nulls, the complexity of $\langle I, T, sup, k, s \rangle$ is NP-complete.*

**Proof.** (*Sketch*) The proof use the same line of reasoning as in Theorem 3.1. However, this time, we use $\epsilon$ values instead of 0 values in the reduction. Furthermore, we note that rules including conditions of the form $I_j \neq 1$ imply that the value of the support is 0, hence only conditions of the form $I_j = 1$ can be taken in account. □

| | $I_1$ | $I_2$ | $I_3$ | $I_4$ | $I_5$ | $I_6$ |
|---|---|---|---|---|---|---|
| $t_{\{v_1,v_2\}}$ | 0 | 0 | 1 | 1 | 1 | 1 |
| $t_{\{v_1,v_4\}}$ | 0 | 1 | 1 | 0 | 1 | 1 |
| $t_{\{v_1,v_5\}}$ | 0 | 1 | 1 | 1 | 0 | 1 |
| $t_{\{v_1,v_6\}}$ | 0 | 1 | 1 | 1 | 1 | 0 |
| $t_{\{v_2,v_3\}}$ | 1 | 0 | 0 | 1 | 1 | 1 |
| $t_{\{v_2,v_4\}}$ | 1 | 0 | 1 | 0 | 1 | 1 |
| $t_{\{v_2,v_5\}}$ | 1 | 0 | 1 | 1 | 0 | 1 |
| $t_{\{v_3,v_4\}}$ | 1 | 1 | 0 | 0 | 1 | 1 |
| $t_{\{v_4,v_5\}}$ | 1 | 1 | 1 | 0 | 0 | 1 |
| $t_{\{v_5,v_6\}}$ | 1 | 1 | 1 | 1 | 0 | 0 |

Figure 1: An example of the reduction used in Theorem 3.1

It is generally believed that when both support and confidence are measured, the latter task (i.e. filtering out rules with low confidence value from a set of rules having support above some threshold) is far easier to compute [3, 20]. We prove next that the problem of finding association rules having high confidence on databases without nulls is a tractable problem, while the same problem on databases with nulls presents *per se* some computational difficulty.

**Lemma 3.1** *Let $I$ be a set of attributes, let $T$ be a database without nulls on $I$, and let $0 < s \leq 1$ be a rational. Then there exists a non trivial association rule $B \Rightarrow H$ on $I$ such that $cnf(B \Rightarrow H, T) \geq s$ iff there exist an attribute $J_H \in I$, a value $u_H \in \mathbf{dom}(J_H, T)$, and a tuple $t \in T$, such that the rule*

$$\left( \bigwedge_{J \in (I - \{J_H\})} (J = t[J]) \right) \Rightarrow (J_H \neq u_H)$$

*is non trivial and has confidence greater than or equal than $s$.*

**Proof.** ($\Rightarrow$) Let

$$(B \Rightarrow H) = (C_1 \wedge \ldots \wedge C_h \Rightarrow C_{h+1} \wedge \ldots \wedge C_k)$$

where $C_i$ is an atomic condition, for each $i = 1, \ldots, k$. Let $J_H = \mathbf{att}(C_k)$, and let $u_H \in (\mathbf{dom}(J_H, T) - \mathbf{dom}(C_k, T))$. Since $C_k$ is non trivial, $u_H$ always exists. Consider the rule

$$(B' \Rightarrow H') = (C_1 \wedge \ldots \wedge C_{k-1} \Rightarrow (J_H \neq u_H))$$

Then

$$cnf(B' \Rightarrow H', T) \geq cnf(B \Rightarrow H, T) \geq s$$

Let $I - \{J_H\} = J_1, \ldots, J_{n-1}$. For each $t \in T$, we denote by $\mathcal{C}(t)$ the condition

$$(J_1 = t[J_1]) \wedge \ldots \wedge (J_{n-1} = t[J_{n-1}])$$

Let $T'$ be a maximal subset of $T_{B'}$ such that for each $t \in T'$ there does not exist $t' \in T' - \{t\}$ such that $(t[J_1] = t'[J_1]) \wedge \ldots \wedge (t[J_{n-1}] = t'[J_{n-1}])$.

We show that there exists $t \in T'$ such that $\frac{|T_{\mathcal{C}(t) \wedge H'}|}{|T_{\mathcal{C}(t)}|} \geq s$. Assume by contradiction, for each $t \in T'$, $\frac{|T_{\mathcal{C}(t) \wedge H'}|}{|T_{\mathcal{C}(t)}|} < s$. Then

> *For each $i = 1, \ldots, |I|$, consider the $i$th attribute $J_i$ of $I$ ;*
>   *Build the ordered database $T^i$ by sorting $T$*
>   *w.r.t. the sequence $J_1, \ldots, J_{i-1}, J_{i+1}, \ldots, J_n, J_i$;*
>     *For each block $B$ of adjacent tuples of $T^i$*
>     *that are identical on the attributes $I - \{J_i\}$;*
>       *Determine the value $b = \min_{u \in \mathbf{dom}(J_i,T)} |\{t \in B \mid t[J_i] = u\}|$;*
>       *If $(|B| - b)/|B| \geq s$ then return "yes";*
> *Return "no";*

Figure 2: The algorithm deciding the confidence problem on databases without nulls

$$cnf(B' \Rightarrow H', T) = \frac{|\bigcup_{t' \in T'} T_{\mathcal{C}(t') \wedge H'}|}{|\bigcup_{t'' \in T'} T_{\mathcal{C}(t'')}|} = \frac{\sum_{t' \in T'} |T_{\mathcal{C}(t') \wedge H'}|}{\sum_{t'' \in T'} |T_{\mathcal{C}(t'')}|} < \frac{\sum_{t' \in D'} s |T_{\mathcal{C}(t')}|}{\sum_{t'' \in T'} |T_{\mathcal{C}(t'')}|} = s$$

But $cnf(B' \Rightarrow H', T) \geq s$. Then there exists $\bar{t} \in T'$ such that $\frac{|T_{\mathcal{C}(\bar{t}) \wedge H'}|}{|T_{\mathcal{C}(\bar{t})}|} \geq s$. Hence $\mathcal{C}(\bar{t}) \Rightarrow H'$ is the required rule. Finally, we note that the rule is clearly non trivial.

($\Leftarrow$) Straightforward. □

**Theorem 3.3** *Given a database $T$ without nulls, the problem $\langle I, T, cnf, k, s \rangle$ is in P.*

**Proof.** (*Sketch*) The problem can be solved in time $\mathcal{O}(|I| \cdot |T|^2 \log |T|)$ by testing if there exists an association rule of the form described in Lemma 3.1, with confidence exceeding the threshold $s$. Figure 2 reports the algorithm deciding the problem $\langle I, T, cnf, k, s \rangle$ on databases without nulls. □

**Proposition 3.3** *Consider the problem $\mathcal{P} = \langle I, T, cnf, k, s \rangle$. If there exists an association rule $B \Rightarrow H$ that is a solution for $\mathcal{P}$, then the rule $B' \Rightarrow H'$ also solves $\mathcal{P}$, where $B' \wedge H' = B \wedge H$ and $|H'| = 1$.*

**Theorem 3.4** *Given a database $T$ with nulls, the complexity of $\langle I, T, cnf, k, s \rangle$ is NP-complete.*

**Proof.** (*Hardness*) The proof, as in Theorem 3.1, is by reduction of CLIQUE. Let $G = (V, E)$ be an undirected graph, with set of nodes $V = \{v_1, \ldots, v_n\}$ and set of edges $E = \{e_1 = \{v_{p_1}, v_{q_1}\}, \ldots, e_m = \{v_{p_m}, v_{q_m}\}\}$. We build an instance of $\langle I, T, cnf, k, s \rangle$ as follows.

Let $I^{clq}$ be $I' \cup \{I_{n+1}\}$, where $I' = \{I_1, \ldots, I_n\}$, $I_j$ represents the node $v_j$ of $G$, for $j = 1, \ldots, n$, and $I_{n+1}$ is a new attribute representing a new node $v_{n+1}$. Let $T^{clq} = T' \cup T''$, where $T'$ includes the tuples $t_{e_i}$ and $t'_{e_i}$, where $t_{e_i}[I_j] = \epsilon$ (resp. $t'_{e_i}[I_j] = \epsilon$) if $v_j \in e_i$, and $t_{e_i}[I_j] = 1$ (resp. $t'_{e_i}[I_j] = 1$) otherwise, for $i = 1, \ldots, m, j = 1, \ldots, n+1$, (the tuples $t_{e_i}$ and $t'_{e_i}$ both denote the edge $e_i$ of $G$).

Furthermore, $T''$ includes the tuples $t_{v_i}$, where $t_{v_i}[I_j] = \epsilon$ if $i = j$, and $t_{v_i}[I_j] = 1$ otherwise, for $i = 1, \ldots, n+1, j = 1, \ldots, n+1$. Next, we prove that there exists a clique of size $k$ in $G$ iff $\langle I^{clq}, T^{clq}, n - k + 1, \frac{k^2}{k^2+1} \rangle$ is a YES instance.

**Fact 3.2** *Let $C$ be a condition on a subset of $I'$, then $|T'_C| \leq 2\binom{n-|C|}{2}$ and $|T''_C| \leq n + 1 - |C|$.*

($\Rightarrow$) Let $C = \{v_{r_1}, \ldots, v_{r_k}\}$ be a clique of size $k$ in $G$. Consider the condition $B = \left(\bigwedge_{v_j \in (V-C)} (I_j = 1)\right)$ such that $|B| = n - k$. By definition of clique, there exist $\frac{k(k-1)}{2}$ edges of $G$ connecting nodes in $C$. Now,

$$T'_B = \{t_{\{v_{r_x}, v_{r_y}\}}, t'_{\{v_{r_x}, v_{r_y}\}} \mid 1 \leq x < y \leq k\}$$

Thus, $|T'_B| = 2\binom{n-|B|}{2} = k(k-1)$, whereas $|T''_B| = n + 1 - |B| = k + 1$. Hence, $|T^{clq}_B| = k(k-1) + (k+1) = k^2 + 1$, and

$$cnf(B \Rightarrow (I_{n+1} = 1), T^{clq}) = \frac{|T^{clq}_{B \wedge (I_{n+1}=1)}|}{|T^{clq}_B|} = \frac{|T^{clq}_B| - 1}{|T^{clq}_B|} = \frac{k^2}{k^2+1}$$

($\Leftarrow$) If $\langle I^{clq}, T^{clq}, n - k + 1, \frac{k^2}{k^2+1}\rangle$ is a YES instance, then there exists $B \Rightarrow H$ on $I^{clq}$, with $|H| = 1$, such that $|B| \geq n - k$ (by Proposition 3.3).
First, we note that the presence in the rule of atomic conditions of the form $I_j \neq 1$ implies that the rule has confidence 0. Hence only atomic conditions of the form $I_j = 1$ can appear in $B \Rightarrow H$.
The content of $T''$ implies that there is no association rule having confidence 1 on $T^{clq}$. Furthermore, we can infer that $|T^{clq}_B| \geq k^2 + 1$, otherwise the ratio $\frac{|T^{clq}_{B \wedge H}|}{|T^{clq}_B|}$ would not be greater than or equal to $\frac{k^2}{k^2+1}$[2]. Two cases are to be considered: $(a)$ $I_{n+1} \notin \mathbf{att}(B)$; $(b)$ $I_{n+1} \in \mathbf{att}(B)$.
(*Case a*) Assume that $\mathbf{att}(B) \subseteq I'$. Then $|T^{clq}_B| \geq k^2 + 1$ implies that $|B| \leq n - k$, and we have already noticed that $|B| \geq n - k$. Thus $|B| = n - k$ and $|T^{clq}_B| = k^2 + 1$. Let $I' - \mathbf{att}(B) = \{I_{r_1}, \ldots, I_{r_k}\}$. Since $|B| = n - k$, then $|T''_B| = k + 1$, whereas, in order to be $|T'_B| = k(k-1)$ it is necessary that

$$T'_B = \{t_{\{v_{r_x}, v_{r_y}\}}, t'_{\{v_{r_x}, v_{r_y}\}} \mid 1 \leq x < y \leq k\}$$

Thus $E \supseteq \{(v_{r_x}, v_{r_y}) \mid 1 \leq x < y \leq k\}$, and the nodes $v_{r_1}, \ldots, v_{r_k}$ form a clique of $G$ having size $k$.
(*Case b*) Suppose that $B = B' \wedge (I_{n+1} = 1)$. Then $|T^{clq}_B| \geq k^2 + 1$ implies that $|B'| \leq n - k - 1$, and we have already noticed that $|B| \geq n - k$, i.e. $|B'| \geq n - k - 1$. Thus $|B'| = n - k - 1$ and (by recalling Fact 3.2)

$$k^2 + 1 \leq |T^{clq}_B| \leq 2\binom{n-|B'|}{2} + (n + 1 - |B'|) = k^2 + 2k + 2$$

We can show that there does not exist a tuple $t \in T'$ such that $t \not\vdash H$ and $t \vdash B$. Assume, by contradiction, that such a tuple $t \in T'$ exists. Then $|T^{clq}_{B \wedge H}| \leq |T^{clq}_B| - 3$. This implies that the confidence of the association rule $B \Rightarrow H$ cannot be greater than or equal to $k^2/(k^2 + 1)$, since

$$(\forall k) \quad \frac{|T^{clq}_B| - 3}{|T^{clq}_B|} \leq \frac{(k^2 + 2k + 2) - 3}{k^2 + 2k + 2} < \frac{k^2}{k^2 + 1}$$

Thus, $H$ is such that $|T'_{B \wedge H}| = |T'_B| = |T'_{B'}| = |T'_{B' \wedge H}|$. Since $|T^{clq}_B| \geq k^2 + 1$ and, by Fact 3.2, we know that $|T^{clq}_{B' \wedge H}| \leq k^2 + 1$ (note that $|B' \wedge H| = n - k$), it follows that $|T^{clq}_{B' \wedge H}| = |T^{clq}_B| = k^2 + 1$. Let $I' - \mathbf{att}(B \wedge H) = \{I_{r_1}, \ldots, I_{r_k}\}$. Hence

$$T'_B = \{t_{\{v_{r_x}, v_{r_y}\}}, t'_{\{v_{r_x}, v_{r_y}\}} \mid 1 \leq x < y \leq k\}$$

Thus $E \supseteq \{(v_{r_x}, v_{r_y}) \mid 1 \leq x < y \leq k\}$, and the nodes $v_{r_1}, \ldots, v_{r_k}$ form a clique of $G$ having size $k$.
(*Membership*) Certificate: an association rule $B \Rightarrow H$ on $I$. Polynomial checking: verify that $B \Rightarrow H$ is non trivial, $|B \wedge H| \geq k$, and $cnf(B \Rightarrow H, T) \geq s$. $\square$

Despite the syntactical similarity with confidence, the laplace metric is closer to support than confidence. Consider the laplace expression. For each rule $B \Rightarrow H$, database $T$, and fixed value of $|T_B|$, laplace is maximum when $|T_{B \wedge H}| = |T_B|$. Assume the above relation is satisfied. In order to be $laplace_h(B \Rightarrow H, T) \geq s$, it must be the case that $|T_{B \wedge H}| \geq \frac{hs-1}{1-s}$. Assume now that $s \to 1$; this implies that $|T_{B \wedge H}| \to \infty$, i.e. that $sup(B \Rightarrow H, T) \to 1$. The following Theorem formalize the above intuitive argumentation.

---

[2] Consider the inequality $\frac{m}{m+1} \geq \frac{m-1}{m}$

|  | $I_1$ | $I_2$ | $I_3$ | $I_4$ | $I_5$ | $I_6$ |
|---|---|---|---|---|---|---|
| $t_{\{v_1,v_2\}}$ | $\epsilon$ | $\epsilon$ | 1 | 1 | 1 | 1 |
| $t'_{\{v_1,v_2\}}$ | $\epsilon$ | $\epsilon$ | 1 | 1 | 1 | 1 |
| $t_{\{v_1,v_3\}}$ | $\epsilon$ | 1 | $\epsilon$ | 1 | 1 | 1 |
| $t'_{\{v_1,v_3\}}$ | $\epsilon$ | 1 | $\epsilon$ | 1 | 1 | 1 |
| $t_{\{v_1,v_4\}}$ | $\epsilon$ | 1 | 1 | $\epsilon$ | 1 | 1 |
| $t'_{\{v_1,v_4\}}$ | $\epsilon$ | 1 | 1 | $\epsilon$ | 1 | 1 |
| $t_{\{v_2,v_3\}}$ | 1 | $\epsilon$ | $\epsilon$ | 1 | 1 | 1 |
| $t'_{\{v_2,v_3\}}$ | 1 | $\epsilon$ | $\epsilon$ | 1 | 1 | 1 |
| $t_{\{v_2,v_4\}}$ | 1 | $\epsilon$ | 1 | $\epsilon$ | 1 | 1 |
| $t'_{\{v_2,v_4\}}$ | 1 | $\epsilon$ | 1 | $\epsilon$ | 1 | 1 |
| $t_{\{v_3,v_4\}}$ | 1 | 1 | $\epsilon$ | $\epsilon$ | 1 | 1 |
| $t'_{\{v_3,v_4\}}$ | 1 | 1 | $\epsilon$ | $\epsilon$ | 1 | 1 |
| $t_{\{v_3,v_5\}}$ | 1 | 1 | $\epsilon$ | 1 | $\epsilon$ | 1 |
| $t'_{\{v_3,v_5\}}$ | 1 | 1 | $\epsilon$ | 1 | $\epsilon$ | 1 |
| $t_{\{v_4,v_5\}}$ | 1 | 1 | 1 | $\epsilon$ | $\epsilon$ | 1 |
| $t'_{\{v_4,v_5\}}$ | 1 | 1 | 1 | $\epsilon$ | $\epsilon$ | 1 |
| $t_{v_1}$ | $\epsilon$ | 1 | 1 | 1 | 1 | 1 |
| $t_{v_2}$ | 1 | $\epsilon$ | 1 | 1 | 1 | 1 |
| $t_{v_3}$ | 1 | 1 | $\epsilon$ | 1 | 1 | 1 |
| $t_{v_4}$ | 1 | 1 | 1 | $\epsilon$ | 1 | 1 |
| $t_{v_5}$ | 1 | 1 | 1 | 1 | $\epsilon$ | 1 |
| $t_{v_6}$ | 1 | 1 | 1 | 1 | 1 | $\epsilon$ |

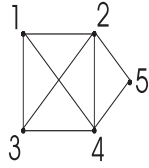

Figure 3: An example of the reduction used in Theorem 3.4

**Theorem 3.5** *Let $T$ be a database without nulls. Then the complexity of $\langle I, T, \rho, k, s \rangle$, with $\rho \in \{gain_\theta, laplace_h\}$, is NP-complete.*

**Proof.** (*Hardness*) Once again, for the hardness part, we use a reduction of CLIQUE. Thus, let $G = (V, E)$ be an undirected graph, with set of nodes $V = \{v_1, \ldots, v_n\}$ and set of edges $E = \{e_1 = \{v_{p_1}, v_{q_1}\}, \ldots, e_m = \{v_{p_m}, v_{q_m}\}\}$. Let $I^{clq}$ be the set of attributes $I_1, \ldots I_n, I_{n+1}$, where $I_j$ denotes the node $v_j$ of $G$ ($j = 1, \ldots, n$) and $I_{n+1}$ is an additional attribute. Furthermore, let $T^{clq}$ include the tuples $t_{e_i}, t'_{e_i}$ s.t. $t_{e_i}[I_j] = t'_{e_i}[I_j] = 0$ if $v_j \in e_i$, and 1 otherwise, where $t_{e_i}$ and $t'_{e_i}$ both denote the edge $e_i$ of $G$, for each $i = 1, \ldots, m$, and the tuple $t_0$, s.t. $t_0[I_j] = 0$ for each $j = 1, \ldots, n+1$.
Let $s^{clq}$ be $\frac{(1-\theta)k(k-1)}{2m+1}$ ($\frac{k(k-1)+1}{k(k-1)+h}$ resp.). Next we prove that there exists a clique of size $k$ in $G$ iff $\langle I^{clq}, T^{clq}, gain_\theta, n-k+1, s^{clq} \rangle$ ($\langle I^{clq}, T^{clq}, laplace_h, n-k+1, s^{clq} \rangle$ resp.) is a YES instance.
We have the following facts.

**Fact 3.3** *Let $I_j \in I^{clq}$, let $C' = (I_j = 0)$ or $C' = (I_j \neq 1)$, and let $C''$ be a non trivial condition on a subset of $I^{clq} - \{I_j\}$. Then $|T^{clq}_{C' \wedge C''}| \leq 2(n - |C' \wedge C''|)$.*

**Fact 3.4** *Let $C$ be a condition on a subset of $I^{clq} - \{I_{n+1}\}$ composed by atomic conditions of the form $I_j = 1$ or $I_j \neq 0$. Then $|T^{clq}_C| \leq 2\binom{n-|C|}{2}$.*

We can resume Theorem's proof.

($\Rightarrow$) Let $C = \{v_{r_1}, \ldots, v_{r_k}\}$ be a clique of size $k$ in $G$. Consider the condition $B = \left(\bigwedge_{v_j \in (V-C)} (I_j = 1)\right)$, and the condition $H = (I_{n+1} = 1)$. Clearly, $B \wedge H$ is non trivial. By definition of clique, there exist $\frac{k(k-1)}{2}$ edges of $G$ connecting nodes in $C$. Therefore, the cardinality of

$$T' = \{t_{\{v_{r_x}, v_{r_y}\}}, t'_{\{v_{r_x}, v_{r_y}\}} \in T^{clq} \mid 1 \leq x < y \leq k\}$$

equals $k(k-1)$. Clearly $T' \subseteq T^{clq}_{B \wedge H}$, and $T^{clq}_{B \wedge H} = T^{clq}_B$, hence

$$gain_\theta(B \Rightarrow H, T^{clq}) \geq s^{clq}$$

($laplace_h(B \Rightarrow H, T^{clq}) \geq s^{clq}$ resp.).

($\Leftarrow$) If $\langle I^{clq}, T^{clq}, gain_\theta, n-k+1, s^{clq}\rangle$ ($\langle I^{clq}, T^{clq}, laplace_h, n-k+1, s^{clq}\rangle$ resp.) is a YES instance then there exists a non trivial rule $B \Rightarrow H$ on $I^{clq}$ such that $|B \wedge H| \geq n-k+1$ and $gain_\theta(B \Rightarrow H, T^{clq}) \geq s^{clq}$ ($laplace_h(B \Rightarrow H, T^{clq}) \geq s^{clq}$ resp.).
We show that only conditions of the form $I_j = 1$ (or $I_j \neq 0$) can appear in $B \wedge H$. First, we note that atomic conditions on numerical attributes of the form $I_j \in [0, 1]$ are trivial, while the same does not apply to categorical attributes. W.l.o.g. suppose $k \geq 3$. By contradiction, suppose that there exists $I_j = 0$ (or $I_j \neq 1$) occurring in $B \wedge H$, then, by Fact 3.3, $|T^{clq}_{B \wedge H}| \leq 2(k-1)$. As $gain_\theta$ ($laplace_h$ resp.) increases when $|T_{B \wedge H}|$ increases and $|T_B|$ decreases, and is maximum for $|T_{B \wedge H}| = |T_B|$, then

$$gain_\theta(B \Rightarrow H, T^{clq}) \leq \frac{(1-\theta)2(k-1)}{2m+1} < s^{clq}$$

($laplace_h(B \Rightarrow H, T^{clq}) \leq \frac{2(k-1)+1}{2(k-1)+h} < s^{clq}$ resp.).
We show that $I_{n+1} \in \mathbf{att}(B \wedge H)$. By contradiction, suppose $I_{n+1} \notin \mathbf{att}(B \wedge H)$. Then, by Fact 3.4, $|T^{clq}_{B \wedge H}| \leq (k-1)(k-2)$, hence

$$gain_\theta(B \wedge H, T^{clq}) \leq \frac{(1-\theta)(k-1)(k-2)}{2m+1} < s^{clq}$$

($laplace_h(B \wedge H, T^{clq}) \leq \frac{(k-1)(k-2)+1}{(k-1)(k-2)+h} < s^{clq}$ resp.).
Let $H' = (I_{n+1} = 1)$. We can obtain from $B \Rightarrow H$ an association rule $B' \Rightarrow H'$ such that $gain_\theta(B' \Rightarrow H', T^{clq}) \geq s^{clq}$ ($laplace_h(B' \Rightarrow H', T^{clq}) \geq s^{clq}$ resp.). Simply take as $B'$ the condition such that $B \wedge H$ equals to $B' \wedge (I_{n+1} = 1)$ (or $B' \wedge (I_{n+1} \neq 0)$). We note that $|B'| \geq n-k$.
As $|T^{clq}_{B' \wedge H'}| = |T^{clq}_{B'}|$, then $gain_\theta(B' \Rightarrow H', T^{clq}) \geq s^{clq}$ ($laplace_h(B \Rightarrow H, T^{clq}) \geq s^{clq}$ resp.) implies that $|T^{clq}_{B'}| \geq k(k-1)$. Thus $|B'| \leq n-k$, and we have already noticed that $|B'| \geq n-k$, then the size of $B'$ is exactly $n-k$.
Let $I^{clq} - \mathbf{att}(B') = \{I_{r_1}, \ldots, I_{r_k}, I_{n+1}\}$. In order to be $|T^{clq}_{B'}| \geq k(k-1)$, $T^{clq}_{B'}$ contains, at least, the set $\{t_{\{v_{r_x}, v_{r_y}\}}, t'_{\{v_{r_x}, v_{r_y}\}} \in T^{clq} \mid 1 \leq x < y \leq k\}$, i.e. the nodes $v_{r_1}, \ldots, v_{r_k}$ form a clique of $G$ having size $k$.

(*Membership*) Certificate: an association rule $B \Rightarrow H$ on a subset of $I$. Polynomial checking: verify that $B \Rightarrow H$ is non trivial, that $|B \Rightarrow H| \geq k$, and that $gain_\theta(B \Rightarrow H, T) \geq s$ ($laplace_h(B \Rightarrow H, T) \geq s$ resp.). $\square$

**Theorem 3.6** *Let $T$ be a database with nulls. Then the complexity of $\langle I, T, \rho, k, s\rangle$, with $\rho \in \{gain_\theta, laplace_h\}$, is NP-complete.*

**Proof.** (*Sketch*) The proof use the same line of reasoning as in Theorem 3.5. However, this time, we use $\epsilon$ values instead of 0 values in the reduction. Furthermore, we note that conditions of the form $I_j \neq 1$ imply that the value of gain and laplace is 0, hence only conditions of the form $I_j = 1$ are admissible. $\square$

```
begin
    for i := 1 to |T| do
        if |t_i| ≥ k then
            for guess := 1 to 2^{|t_i|} − 1 do
                if guess has exactly k bits set to 1 then begin
                    count := 0;
                    for j := 1 to |T| do
                        if SATISFIES(t_j, guess, t_i) then count := count + 1;
                    if count ≥ s|T| then return "yes";
                end;
    return "no";
end.

function SATISFIES(v, guess, u) : boolean;
begin
    p := 1;
    for q := 1 to |I| do
        if u[A_q] = c(A_q) then begin
            if guess[p] = 1 and v[A_q] = ϵ then return false;
            p := p + 1;
        end;
    return true;
end; { SATISFIES }
```

Figure 4: The algorithm of Theorem 4.1

## 4 Sparse databases

There are many real applications having associated sparse databases. As an example consider a database of transactions from a large market stored for basket analysis purpose. For databases showing this property, complexity figures are quite different from what we have proved above.

**Theorem 4.1** *Let $T$ be a sparse database. Then the complexity of $\langle I, T, sup, k, s \rangle$ is in $L$.*

**Proof.** We can build a Turing Machine $\mathcal{T}$ employing $\mathcal{O}(\log(\max\{|I|, |T|\}))$ space, which decides $\langle I, T, sup, k, s \rangle$.

Let $T = \{t_1, \ldots, t_m\}$, and let $I = \{A_1, \ldots, A_n\}$. Let $guess$ be a (log-space) counter, and let $p$ be an integer, then $guess[p]$ denotes the value of the $p$-th bit of $guess$. The algorithm which is followed by $\mathcal{T}$ is depicted in Figure 4.

Roughly speaking, $\mathcal{T}$ considers each tuple $t_i$, using the counter $i$, and tests only those conditions which can be built on $t_i$. It is not necessary to represent each condition explicitly; the counter $guess$ is employed instead: the $p$-th bit of $guess$ tells whether the $p$-th non null attribute value occurring in $t_i$ belongs to the current condition or not. Each guessed condition is then tested on each transaction $t_j$ of $T$, using the counter $j$. The counter $count$ takes into account the number of tuples satisfying the current condition.

It is straightforward to note that the space employed corresponds to the space needed to store the variables $i, j, count, p, q$ and $guess$. On the assumption that $T$ is sparse, $i, j$ and $count$ need $\mathcal{O}(\log |T|)$ space, whereas $p, q$ and $guess$ need $\mathcal{O}(\log |I|)$ space. Finally, verifying if $guess$ has at least $k$ bits set to 1 can be easily done in logarithmic space.  □

**Theorem 4.2** *Let $T$ be a sparse database. Then the complexity of $\langle I, T, \rho, k, s \rangle$, where $\rho \in \{cnf, gain_\theta, laplace_h\}$ is in $L$.*

**Proof.** (*Sketch*). The proof follows the same line of reasoning of Theorem 4.1. In this case, two disjoint current conditions are needed (which represent the body and the head of the current association rule, respectively), and some further auxiliary counters using logarithmic space. □

## 5 Fixed schema complexity

In this Section we improve the result reported in [19], stating the polynomial-time solvability of the association rule mining problem under the fixed schema complexity measure. For simplicity, we give only the proof regarding numerical attributes.

**Theorem 5.1** *Let $I$ be a set of numerical attributes. Then the fixed schema complexity of the problem $\langle I, T, sup, k, s \rangle$ is in L.*

**Proof.** Let $n = |I|$, and let $m = |T|$. We can build a Turing Machine $\mathcal{T}$ employing $\mathcal{O}(\log m)$ space, which solves $\langle I, T, sup, k, s \rangle$. $\mathcal{T}$ use $2n$ pointers $p_j^l$, $p_j^u$, to $2n$ tuples of $T$, of size $\mathcal{O}(\log m)$ each, and $2n$ bits $o_j$ and $i_j$, for each $j = 1, \ldots, n$. An *arrangement* of $\mathcal{T}$ is a $4n$-tuple $(p_1^l, p_1^u, \ldots, p_n^l, p_n^u, o_1, \ldots, o_n, i_1, \ldots, i_n) \in \{1, \ldots, m\}^{2n} \times \{0, 1\}^{2n}$. Let $t_i$ denote the $i$-th tuple of $T$; define $\theta(0)$ as "$\in$", $\theta(1)$ as "$\notin$", and $C_j$ as $I_j \ \theta(o_j) \ [t_{p_j^l}[I_j], t_{p_j^u}[I_j]]$, for each $j = 1, \ldots, n$.

$\mathcal{T}$ works as follows: it scans, one after one, all the arrangements, and for each of them performs the following steps: (1) Verifies that $i_1 + \ldots + i_n = k$; (2) If step 1 succeeds, verifies that $t_{p_j^l}[I_j] \neq \epsilon$ and $t_{p_j^u}[I_j] \neq \epsilon$, for each $j = 1, \ldots, n$ such that $i_j = 1$; (3) If step 2 succeeds, verifies that $t_{p_j^l}[I_j] \leq t_{p_j^u}[I_j]$, for each $j = 1, \ldots, n$ such that $i_j = 1$; (4) If step 3 succeeds, verifies that the conditions $C_j$, for each $j = 1, \ldots, n$ such that $i_j = 1$, are non trivial; (5) If step 4 succeeds, verifies that $|T_{\bigwedge_{i_j=1} C_j}| \geq s|T|$; (6) If step 5 succeeds, return "yes" and stops.

If $\mathcal{T}$ does not reach step 5, finally return "no" and stops. We note that, to execute steps 1-5, the Turing Machine needs an additional amount of space, to store counters and auxiliary pointers, that is logarithmic w.r.t. the input size. It follows that $\mathcal{T}$ returns "yes" iff $\langle I, T, sup, k, s \rangle$ is a YES instance. □

**Theorem 5.2** *The fixed schema complexity of the problems $\langle I, T, \rho, k, s \rangle$, where $\rho \in \{cnf, gain_\theta, laplace_h\}$, is in L.*

**Proof.** (*Sketch*) The proof use the same line of reasoning as in Theorem 5.1. □

## 6 Further complexity results

In this section, we investigate the computational complexity of several interesting special cases of mining association rules. Most of them assume some parameters (e.g., the lower bound on the rule length $k$, the index value threshold $s$) of the general association rule mining problem to be fixed. The relevance of the analysis we present below is two-fold. First, it eases the task of detecting actual complexity sources. Second, from a practical point of view, users are often interested in solving such simplified tasks, as, for instance, when one wishes to mine only rules with a support always larger than 75 percent.

As stated below, the rule mining problem remains very hard to solve whenever the support threshold is kept fixed.

**Theorem 6.1** *The problem $\langle I, T, sup, k, s \rangle$ where $s$ is a fixed constant in $(0, 1)$, and $T$ is a database with nulls is $NP$-complete.*

**Proof.** Let $I$ be a set of attributes $I_1, \ldots, I_n$ defined on the domain $\{\epsilon, c\}$. Let $T$ be a boolean database defined on $I$ and let $S$ be a subset of $I$. A tuple $t$ s.t. $t[J] = c$, for each $J \in I$, and s.t. $t[J] = \epsilon$ otherwise, will be defined in the following as $t = S$.

(*Hardness*) The proof is by reduction of CLIQUE. Let $G = (V, E)$ be an undirected graph, with set of nodes $V = \{v_1, \ldots, v_n\}$ and set of edges $E = \{(v_{p_1}, v_{q_1}), \ldots, (v_{p_m}, v_{q_m})\}$. We build a corresponding instance of $\langle I, T, sup, k, s \rangle$ as follows: let $I^{clq}$ be the set consisting of the attributes $I_1, \ldots I_n, I_{n+1}$, where $I_j$ represents the node $v_j$ of $G$, for $j = 1, \ldots, n$ and $I_{n+1}$ is an additional attribute. Let $T^{clq}$ be a set composed by the union of the following sets of tuples:

- $T^G$, including the tuples $t_i = I^{clq} - \{I_{p_i}, I_{q_i}, I_{n+1}\}$, where $t_i$ represents the edge $(v_{p_i}, v_{q_i})$ of $G$, for each $i = 1, \ldots, m$;
- $T^0$, including $c_0$ copies of the tuple $\{I_{n+1}\}$, where $c_0$ is a value to be defined next;
- $T^1$, consisting of $c_1$ copies of the tuples $I^{clq} - \{I_{n+1}\}$, where $c_1$ is a value to be defined next.

As for the values $c_0$ and $c_1$ we choose two positive or null integer values such that

$$s = \frac{\frac{k(k-1)}{2} + c_1}{m + c_0 + c_1}$$

It can be shown that such two values exist, and are both polynomial bounded in $m$. Indeed, let $\alpha = k(k-1)/2$, and $s = ax/(bx)$: we have

$$\frac{ax}{bx} = \frac{\alpha + c_1}{m + c_0 + c_1}$$

where $a, b$ and $x$ are positive integers and $a < b$. Thus, $c_0 = ax - \alpha$ and $c_1 = bx - m - (ax - \alpha)$. Setting $x$ equal to, e.g., $m + \alpha$, yields the two required values.

Next, we prove that there exists a clique of size $k$ in $G$ iff $\langle I^{clq}, T^{clq}, sup, n-k, s \rangle$ is a YES instance.

($\Rightarrow$) Let $C = \{v_{r_1}, \ldots, v_{r_k}\}$ be a clique of size $k$ in $G$. Consider the condition

$$B \wedge H = \left( \bigwedge_{v_j \in (V-C)} (I_j = 1) \right)$$

By definition of clique, there exist $k(k-1)/2$ edges of $G$ connecting nodes in $C$, i.e. we can build a set $T' = \{(I^{clq} - \{I_{r_x}, I_{r_y}, I_{n+1}\}) \in T^G \mid 1 \leq x < y \leq k\}$ of $k(k-1)/2$ tuples. Clearly, $T' \subseteq T^{clq}_{B \wedge H}$. Thus $|T^{clq}_{B \wedge H}| \geq k(k-1)/2 + c_1$ and $sup(B \wedge H, T) \geq s$.

($\Leftarrow$) W.l.o.g. suppose $n-k \geq 2$. By Proposition 3.2, if $\langle I^{clq}, T^{clq}, sup, n-k, s \rangle$ is a YES instance then there exists a rule $B \Rightarrow H$ of length $n-k$ and s.t. $|T^{clq}_{B \wedge H}| \geq k(k-1)/2 + c_1$. Since $n - k \geq 2$, $B \wedge H$ cannot contain a condition $I_{n+1} = 1$. We have, indeed, that $\forall J \in I^{clq} : J \neq I_{n+1}$ then $|T^{clq}_{J=1 \wedge I_{n+1}=1}| = \emptyset$. Let $Z = B \wedge H$.

Note that each transaction in $T^G$ has size $n - 3$ and no duplicate item exists. In order to be $|T^{clq}_Z| \geq k(k-1)/2 + c_1$, $T^{clq}_Z$ contains, at least, the set

$$T' = \{(I^{clq} - \{I_{r_x}, I_{r_y}, I_{n+1}\} \in T^G \mid 1 \leq x < y \leq k\}$$

i.e. the nodes $v_{r_1}, \ldots, v_{r_k}$ form a clique of $G$ having size $k$.

(*Membership*) Certificate: a condition $C$. Polynomial checking: verify that $|C| \geq k$ and $sup(C, T) \geq s$. □

Note that the special case $\langle I, T, sup, k, 1 \rangle$ can be easily shown to be in P.

16**Lemma 6.1** *Let C be a condition on a set of boolean attributes. Then there exists a family $\{count(C)_{m,n}\}$ [3] of $\#AC_2^0$ circuits computing $|T_C|$ over any input database T defined on a set of boolean attributes I such that $\mathbf{att}(C) \subseteq I$.*

**Proof.** Let $\mathbf{att}(C) \subseteq I = \{A_1, \ldots, A_n\}$. We define the family $\{count(C)_{m,n}\}$ of $\#AC_2^0$ circuits as follows. The circuit $count(C)_{m,n}$ has $m \times n$ binary inputs $x_{i,j}$, $i = 1, \ldots, m$, $j = 1, \ldots, n$, with $m = |T|$ and $n = |I|$. The input $x_{i,j}$ is 1 if $t_i[A_j] = c(A_j)$, 0 otherwise (i.e. if $t_i[A_j] = \epsilon$). The first level of $count(C)_{m,n}$ consists of $m$ $\times$-gates $G_i$, for $i = 1, \ldots, m$. Each gate $G_i$ receives the $|C|$ inputs $\{x_{i,k} \mid A_k \in \mathbf{att}(C)\}$. Thus the output of $G_i$ is 1 iff $t_i \vdash C$. The second level of $count(C)_{m,n}$ consists of a single $+$-gate receiving in input the outputs of all the $G_i$ gates, for $i = 1, \ldots, m$. Thus the circuit $count(C)_{m,n}$ calculates $|T_C|$ when the input has size $m \times n$. □

The forthcoming Theorems (6.2, 6.3 and 6.4) associate some task related to mining association rules to very low complexity classes such as $TC^0$ and $AC^0$. It turns out that these problems are highly parallelizable (recall that $AC^0 \subset TC^0 \subseteq NC^1$, [12]).

**Theorem 6.2** *Let I be a set of boolean attributes, and let k be a fixed constant. Then the complexity of $\langle I, T, sup, k, s \rangle$ is in $TC^0$.*

**Proof.** Let $s$ be codified as a pair of naturals $(a, b)$ such that $s = a/b$, and let $C$ be a condition on a subset of $I$. Consider the function $f(C, T, s) = (b|T_C| + 1) - a|T|$ taking value over integers. Let $B \Rightarrow H$ be an association rule on $I$, and let $I_R$ be the set $\mathbf{att}(B \wedge H)$. Clearly, $sup(B \Rightarrow H, T) \geq s$ iff $f(B \wedge H, T, s) > 0$.

We recall the following result [5]: for each integer $N$ there exists a log-time uniform $\#AC^0$ circuit, which computes $N$, when the binary representation of $N$ is given in input. Say this circuit $number(N)$. Since $a$ and $b$ are integers, we can build two $\#AC^0$ circuits computing the functions $b|T_C|$ and $a|T| = am$, connecting $number(b)$ to $count(C)_{m,n}$ and $number(a)$ to $number(m)$ through a $\times$-gate, respectively.

Then, the function $f(C, T, s)$ is in the class $GapAC^0$, and the language $\{B \Rightarrow H \text{ on } I \mid sup(B \Rightarrow H, T) \geq s\}$ is in the class $PAC^0$ which coincides with $TC^0$ under log-space uniformity [1, 5]. Thus, there exists a constant-depth polynomial size uniform family $\{C'(I_R)_{m,n}\}$ of circuits of unbounded fan-in AND, OR and MAJORITY gates, such that $C'(I_R)_{m,n}$ outputs 1 iff $sup(B \Rightarrow H, T) \geq s$, when the input database has size $m \times n$. We can build a $TC^0$ family circuits solving the $\langle I, T, sup, k, s \rangle$ problem when $k$ is fixed as follows. Consider the circuit $C(I)_{m,n}$ obtained connecting the outputs of all the circuits $C'(I_R)_{m,n}$, with $I_R \subseteq I$ such that $|I_R| = k$ through an OR gate. Since the number of these circuits is $\binom{|I|}{k} = \mathcal{O}(|I|^k)$, hence polynomial, $C_{m,n}(I)$ has constant depth and polynomial size as well. The result then follows from Proposition 3.2. □

It is of interest to investigate the complexity of mining association rules when the value $s|T|$ is fixed. In this case $\langle I, T, sup, k, s \rangle$ corresponds to the problem of finding an association rule satisfied by almost a fixed number of transactions. Such a problem becomes of relevance when it is necessary to find a fixed size set of transactions satisfying a certain property (e.g. in statistic sampling, see [18]).

**Definition 6.1** *Given a set of boolean attributes $I = \{A_1, \ldots, A_n\}$, and a database $T = \{t_1, \ldots, t_m\}$ defined on $I$, we define $\langle I, T \rangle^{-1}$ to be equal to the pair $\langle I', T' \rangle$, where $I' = \{A'_1, \ldots, A'_m\}$ is a set of boolean attributes, where each $A'_j$ denotes the jth tuple of $T$, for $j = 1, \ldots, m$, and $T' = \{t'_1, \ldots, t'_n\}$ is a database defined on $I'$, with $t'_i$ such that $t'_i[A'_j] = 1$ if $t_j[A_i] = c(A_i)$, and $t'_i[A'_j] = \epsilon$ otherwise (i.e. if $t_j[A_i] = \epsilon$), corresponding to the ith attribute of $I$, for $i = 1, \ldots, n, j = 1, \ldots, m$.*

---
[3] Note that here and elsewhere, by little abuse of notation, for simplicity, we denote a circuit family recognizing inputs in the form of a $m \times n$ boolean matrix by using the subscript $m, n$ instead of one single subscript specification denoting the input size.

**Proposition 6.1** Let be $I$ a set of boolean attributes, let $T$ be a database on $I$, let $k$ be a natural number, $1 \leq k \leq |I|$, let $s$, $0 \leq s \leq 1$, be a rational number, and let $\langle I', T' \rangle = \langle I, T \rangle^{-1}$. Then:

$\langle I, T, sup, k, s \rangle$ is a YES instance $\iff$

$$\langle I', T', sup, \lceil s|T| \rceil, \frac{k}{|I|} \rangle \text{ is a YES instance} \quad (1)$$

**Proof.** $\langle I, T, sup, k, s \rangle$ is a YES instance iff there exist an association rule $B \Rightarrow H$ on $I$ s.t. $|B \Rightarrow H| \geq k$, and $|T_{B \wedge H}| \geq \lceil s|T| \rceil$ iff there exist an association rule $B' \Rightarrow H'$ on $I'$ s.t. $|B' \Rightarrow H'| \geq \lceil s|T| \rceil$ and $|T'_{B' \wedge H'}| \geq k$ iff $\langle I', T', sup, \lceil s|T| \rceil, \frac{k}{|I|} \rangle$ is a YES instance. $\square$

**Theorem 6.3** *Let $I$ be a set of boolean attributes, and let $\lceil s|T| \rceil$ be a fixed constant. Then the complexity of $\langle I, T, sup, k, s \rangle$ is in $TC^0$.*

**Proof.** The result follows immediately from Theorem 6.2 and Proposition 6.1. $\square$

**Theorem 6.4** *Let $I$ be a set of boolean attributes, and let $k$ and $\lceil s|T| \rceil$ two fixed constants. Then the complexity of $\langle I, T, sup, k, s \rangle$ is in $AC_2^0$.*

**Proof.** Let $I = \{A_1, \ldots, A_n\}$, and let $T = \{t_1, \ldots, t_m\}$. Let $B \Rightarrow H$ be an association rule on $I$, and let $I_R$ be the set $\mathbf{att}(B \wedge H)$. Define the family $\{C'(I_R)_{m,n}\}$ of $AC_3^0$ circuits as follows.

The circuit $C'(I_R)_{m,n}$ has $n \times m$ binary inputs $x_{i,j}$, $i = 1, \ldots, m$, $j = 1, \ldots, n$, with $m = |T|$ and $n = |I|$. The input $x_{i,j}$ is 1 if $t_i[A_j] = c(A_j)$, 0 otherwise (i.e. if $t_i[A_j] = \epsilon$). The first level of $C'(I_R)_{m,n}$ consists of $m$ AND gates $G_i^1$, for $i = 1, \ldots, m$. Each gate $G_i^1$ receives the $|I_R|$ inputs $\{x_{i,k} \mid A_k \in I_R\}$.

Thus the output of $G_i^1$ is 1 iff $t_i \vdash (B \wedge H)$. The second level of $C'(I_R)_{m,n}$ consists of $\binom{m}{\lceil sm \rceil}$ AND gates $G_j^2$, for $j = 1, \ldots, |g|$ where

$$g = \{F \subseteq \{G_1^1, \ldots, G_m^1\} : |F| = \lceil sm \rceil\}$$

The gate $G_j^2$ receives in input the outputs of the $\lceil sm \rceil$ gates contained within the $j$-th element of $g$.

The third level consists of a single OR gate receiving in input the outputs of all the $G_j^2$ gates, for $j = 1, \ldots, \binom{m}{\lceil sm \rceil}$. Thus the circuit $C'(I_R)_{m,n}$ decides if $|T_{B \wedge H}| \geq \lceil sm \rceil$. The size of each circuit $C'(I_R)_{m,n}$ is polynomial, since $|g| \leq m^{\lceil sm \rceil}$, and $\lceil sm \rceil$ is fixed. We can build an $AC^0$ circuit solving $\langle I, T, sup, k, s \rangle$, for $k$ and $\lceil s|T| \rceil$ fixed, as follows. Consider the circuit $C(I)_{m,n}$ obtained connecting the outputs of all the circuits $C'(I_R)_{m,n}$, with $I_R \subseteq I$ such that $|I_R| = k$ (it suffices from Proposition 3.2), through an OR gate.

Since the number of these circuits is $\binom{|I|}{k} = \mathcal{O}(|I|^k)$, hence polynomial, $C_{m,n}(I)$ has constant depth and polynomial size as well. The first and second level (of AND gates), and the third and fourth level (of OR gates), can be easily each reorganized into a single level, thus giving an overall circuit family of depth 2. Hence the result follows. $\square$

## 7 Conclusions

In this paper, we have analyzed the computational complexity of mining association rules. We have considered the most widely accepted form of association rules that use well-known quality indices, namely, support, confidence, gain and laplace. After having formally defined association rule mining problems, we have shown that the general versions of these problems are NP-complete, except when confidence is measured on database without nulls.

Then, we have focused on analyzing several interesting restricted cases, for most of which lower complexity bounds have been proved to hold. It is relevant to note that these cases are often related to complexity classes for which the existence of highly parallelizable algorithms has been proved. For example, for sparse databases, the complexities of the mining problem lies within L. In some other analyzed cases, where some of the parameters of the mining problems are considered as fixed constants, the mining problem lies in $TC^0$ or in $AC^0$.

The complexity analysis presented in this paper is not complete, though. For instance, it is relevant to analyze the complexity induced by adopting other indices as, for instance, *entropy* and *improvement* [14, 13]. Moreover, other forms of association rules could be considered as, for instance, sequential patterns [4]. We leave these topics to future research.

| $\rho$ | Database Type | Constraint | Complexity | Reference |
|---|---|---|---|---|
| $sup, gain_\theta, laplace_h$ | no nulls | | NP-complete | Th. 3.1, 3.5 |
| $cnf$ | no nulls | | P | Th. 3.3 |
| all | with nulls | | NP-complete | Th. 3.2, 3.4, 3.6 |
| all | sparse | | L | Th. 4.1, 4.2 |
| all | no nulls | $|I|$ fixed | L | Th. 5.1, 5.2 |
| $sup$ | with nulls | $s$ fixed | NP-complete | Th. 6.1 |
| $sup$ | boolean | $k$ fixed | $TC^0$ | Th. 6.2 |
| $sup$ | boolean | $s|T|$ fixed | $TC^0$ | Th. 6.3 |
| $sup$ | boolean | $s|T|$ and $k$ fixed | $AC_2^0$ | Th. 6.4 |

Figure 5: Summary of complexity results for $\langle I, T, \rho, k, s \rangle$.